\documentclass[twocolumn,prb]{revtex4}
\usepackage{graphicx}
\usepackage{bm}
\usepackage{amsmath}
\begin{document} \title{Branch-entangled polariton pairs in
planar microcavities and photonic wires}

\author{C. Ciuti}
\affiliation{Laboratoire Pierre Aigrain, Ecole Normale
Sup\'erieure, 24, rue Lhomond, 75005 Paris, France }

\begin{abstract} A scheme is proposed for the generation of
branch-entangled pairs of microcavity polaritons through
spontaneous inter-branch parametric scattering.
Branch-entanglement is achievable when there are two twin
processes, where the role of signal and idler can be exchanged
between two different polariton branches. Branch-entanglement of
polariton pairs can lead to the emission of frequency-entangled
photon pairs out of the microcavity. In planar microcavities, the
necessary phase-matching conditions are fulfilled for pumping of
the upper polariton branch at an arbitrary in-plane wave-vector.
The important role of nonlinear losses due to pair scattering into
high-momentum exciton states is evaluated. The results show that
the lack of protection of the pump polaritons in the upper branch
is critical. In photonic wires, branch-entanglement of
one-dimensional polaritons is achievable when the pump excites a
lower polariton sub-branch at normal incidence, providing
protection from the exciton reservoir.
\end{abstract}

\date{\today} \maketitle
%\draft

The generation of entangled states is one of the most fascinating
aspects of quantum mechanics\cite{Haroche}. In quantum optics,
parametric sources of entangled photon pairs have been attracting
great interest due to their remarkable non-classical applications.
In particular, polarization-entangled pairs of
photons\cite{EPRsource} are an essential ingredient for quantum
cryptography \cite{Gisin}, while frequency-entangled pairs have
been recently exploited for the so-called quantum optical
coherence tomography \cite{tomography}. In atomic physics,
parametric collisions are also enjoying considerable attention
with the possibility of creating entangled pairs of atoms by
parametric scattering off a Bose-Einstein
condensate\cite{Meystre}.

Recently, semiconductor quantum microcavities in the strong
exciton-photon coupling regime\cite{Weisbuch,Houdre} have been
shown to provide very rich parametric phenomena
\cite{review,Savvidis_00,Savvidis_Rapid,Stevenson,Baumberg,Messin,Saba,Huynh,Kundermann}.
In these systems, the strong coupling between quantum well exciton
and cavity photon modes gives rise to two branches of quasi-two
dimensional bosons, the so-called lower and upper branch
polaritons. In a polariton device, the parametric scattering is
due to polariton-polariton interactions, which are extremely
efficient\cite{Savvidis_00,Saba}. Moreover, the energy-momentum
conservation (phase-matching) can be provided intrinsically by the
peculiar shape of the polariton energy dispersion. Interestingly,
semiconductor planar microcavities can be laterally patterned with
the possibility of creating zero-dimensional\cite{Dasbach} and
one-dimensional\cite{Dasbach_wire} polariton systems with
controllable parametric properties. Efficient inter-branch
parametric scattering has been demonstrated in one-dimensional
microcavities\cite{Dasbach_wire}, where the presence of several
polariton sub-branches provides the opportunity of tailoring the
parametric processes in a remarkable way.

While the outstanding optical gain properties of polariton
parametric amplifiers involving the lower branch are largely
investigated, the study of the genuine quantum properties is still
in its infancy. So far, current research has been focused on the
generation and detection of polariton squeezing\cite{Schwendimann}
due to the anomalous correlation between signal and idler
polaritons, both belonging to the lower branch. Polariton
squeezing has been recently demonstrated in the degenerate
configuration\cite{Messin,squeezing} (signal, pump and idler in
the same lower polariton branch mode), but the detection of
two-mode squeezing in the non-degenerate configuration appears
challenging due to the very different extra-cavity radiative
coupling of signal and idler modes within the lower
branch\cite{twin}.

One important issue yet to be explored is the possibility of
creating Einstein-Podolski-Rosen (EPR) pairs of polaritons, which
are entangled with respect to a certain degree of freedom and
which can be efficiently transferred out of the microcavity. In
this paper, we propose a scheme to create polariton pairs, which
are entangled with respect to a peculiar degree of freedom, namely
the discrete polariton branch index. We show that spontaneous
inter-branch parametric scattering can generate pairs in the
entangled state of the form
\begin{equation}
|\Psi \rangle \propto
  |j_1,k_{\text{s}}\rangle
   |j_2,k_{\text{i}}\rangle
 +~
  |j_2,k_{\text{s}}\rangle
  |j_1,k_{\text{i}}\rangle
 ~,
 \end{equation}
where $|j,k\rangle$ denotes a polariton state belonging to the
$j$-th branch (or sub-branch) mode with wave-vector $k$. The
signal and idler wave-vectors ($k_{\text{s}}$ and $k_{\text{i}}$)
are such to provide phase-matching for the two branch-exchanged
processes, as it will be discussed later in detail. We show that
the necessary (but not sufficient) phase-matching requirements for
this kind of parametric effect are easily fulfilled both in
two-dimensional systems (planar microcavities) and one-dimensional
structures (photonic wires), thanks to the dispersion of polariton
branches, which can be engineered. In our study, we evaluate the
protection of the considered parametric process from nonlinear
losses (collision broadening). In planar microcavities, we find
that pair scattering into the exciton reservoir can be a severe
limitation. In fact, when the pump drives the upper branch,
pump-pump, pump-signal and pump-idler scattering into the
high-momentum exciton states is particularly efficient. In
photonic wires, this lack of protection of pump polaritons in the
upper branch can be naturally defeated. In fact, in photonic
wires, the additional confinement of the photon modes produces a
many-fold of sub-branches. In these systems, inter-branch
scattering is possible even under pump excitation of the lower
sub-branch, as recently demonstrated experimentally
\cite{Dasbach_wire}. We show that by pumping a lower sub-branch at
normal incidence ($k_x = 0$), branch-entangled pairs of polaritons
with a finite wave-vector can be obtained. Since the pumped mode
lies in a lower sub-branch, pump-pump, pump-signal and pump-idler
scattering into the exciton reservoir can be suppressed.

The paper is organized as follows. In Section \ref{2D}, we
describe the proposed inter-branch process in a planar
microcavity, where the upper polariton branch is excited. The
generation of branch-entangled pairs of polaritons is treated
within a quantum Hamiltonian model, presented in Section
\ref{model}. Section \ref{out} treats the coupling to the
extra-cavity field, which is responsible for the spontaneous
emission of frequency-entangled pairs of photons. In Section
\ref{losses} and \ref{threshold}, we address the important issue
of nonlinear losses. In Section \ref{wire}, we consider the case
of photonic wires. Finally, conclusions are drawn in Section
\ref{conclusion}.
\section{2D microcavities}
\subsection{Phase-matching for inter-branch scattering}
\label{2D}

\begin{figure}[t]
\includegraphics[width=7.5cm]{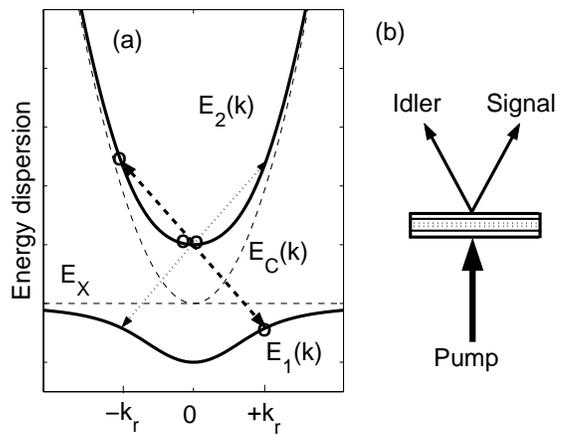}
\caption{(a) Solid lines: in-plane energy dispersion $E_1(k)$
($E_2(k)$) for the lower (upper) polariton branch. Dashed lines:
dispersion $E_C(k)$ ($E_X(k)$) of the cavity (exciton) mode.
Arrows depict the considered inter-branch polariton pair
scattering process. (b) Sketch of the excitation geometry of the
planar microcavity. \label{fig_00kk_disp}}
\end{figure}

\begin{figure}
\includegraphics[width=8cm]{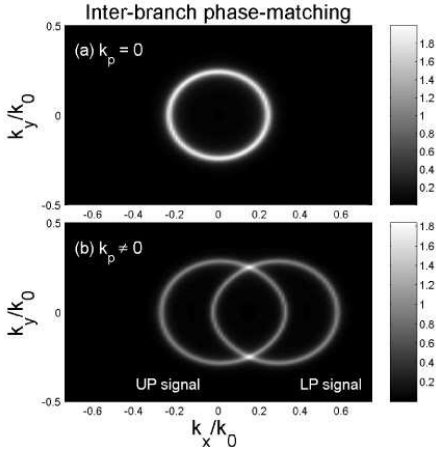}
\caption{Phase-matching function $\eta({\bf k})$ (defined in the
text) as a function of the signal in-plane wave-vector ${\bf k}$
($k_0$ units). (a) The pump excites the upper branch at normal
incidence (${\bf k_{\text{p}}} = {\bf 0}$). (b) ${\bf
k_{\text{p}}} = 0.15~{k_0}~\hat{\bf x}$.  Parameters: $E_X =
E_C(0) = 1.5~e\text{V}$, $k_0 = E_C(0)/(\hbar c)$, $2 \hbar
\Omega_\text{R} = 4 ~\text{m}e\text{V}$, $\gamma = 0.5
~\text{m}e\text{V}$.
 \label{fig_circles}}
\end{figure}

We start by giving the general idea of the proposed process and
then we turn to a more detailed theoretical analysis. The strong
coupling between exciton and cavity photon modes is known to
produce an anticrossing of their energy dispersions $E_{C}(k)$ and
$E_{X}(k)$, resulting in the appearance of the lower and upper
polariton branches, whose energy dispersions $E_{1}(k)$ and
$E_{2}(k)$ are depicted in in Fig. \ref{fig_00kk_disp}(a). So far,
studies of polariton parametric scattering in planar microcavities
have focused on the lower branch, in particular under pump
excitation near the inflection point of the lower branch
dispersion. Here, we consider a different process, which involves
both branches. Suppose a pump laser injects polaritons in the
upper branch state with zero in-plane wave-vector (${\bf
k_{\text{p}}} = {\bf 0}$). Two injected upper polaritons can
scatter coherently, being parametrically converted into a
signal-idler pair of polaritons, namely a lower and an upper
polariton with opposite in-plane momentum (see
Fig.\ref{fig_00kk_disp}(a) and (b)). The phase-matching is
fulfilled when the idler and signal wave-vector are such that
$|{\bf k_{\text{s}}}| = |{\bf k_{\text{i}}}| = k_{\text{r}}$,
where $k_{\text{r}}$ depends on the polariton splitting and
exciton-photon detuning. Note that for a given ${\bf
k_{\text{s}}}$, there are two equivalent processes, where the role
of signal and idler is exchanged between the lower and upper
polariton branch. Quantum entanglement is due to our ignorance on
which of the two scattered polaritons is in the lower or upper
branch. Fig. \ref{fig_circles} depicts the phase-matching pattern
in the two-dimensional momentum space. We have plotted the
phase-matching function $\eta({\bf k}) = \eta_1({\bf k}) +
\eta_2({\bf k})$, with
\begin{equation}
\eta_{1(2)}({\bf k})= \frac{\gamma^2}{(E_{1(2)}({\bf
k})+E_{2(1)}(2{\bf k_p}-{\bf k})-2E_{2}({\bf k_p}))^2 +
\gamma^2}~,
\end{equation}
where $\gamma$ represents the polariton broadening. Note that if
the energy-momentum conservation for the inter-branch scattering
is strongly violated, $\eta_{1(2)}({\bf k}) \to 0$. On the other
hand, when $\bf k$ is an exact phase-matching wave-vector for a
lower (upper) polariton signal, $\eta_{1(2)}({\bf k}) = 1 $.
Importantly, if a wave-vector $\bf k$ is phase-matching for both
branches, then $\eta({\bf k}) = 2 $. Fig. \ref{fig_circles}(a)
shows the case ${\bf k_p} = {\bf 0}$, where $\eta_{1}({\bf k}) =
\eta_{2}({\bf k})$ and $\eta({\bf k}) = 2$ on the ring $|{\bf k}|
= k_{\text{r}}$. Entangled polariton pairs can be achieved with
opposite momentum on the ring. On the other hand, Fig.
\ref{fig_circles}(b) shows the case ${\bf k_p} \neq {\bf 0}$,
where the lower and upper branch signal phase-matching curves
split ($\eta_{1}({\bf k}) \neq \eta_{2}({\bf k})$) and
branch-entanglement is possible only at the two intersection
points. Note that this phase-matching profile is topologically
different from the $\infty$-shaped profile obtained under pumping
of the lower branch\cite{Ciuti_01}. Moreover, we point out that
the pattern in Fig. \ref{fig_circles}(b) is reminiscent of the one
achieved in type-II parametric down-conversion, which generates
polarization-entanglement of photon pairs\cite{EPRsource}.

\subsection{Quantum Hamiltonian description}
\label{model} We now turn to a detailed treatment of this system.
As a result of the strong exciton-photon coupling, the lower and
upper polariton boson operators $p_{1,\bf k }$ and $p_{2,\bf k }$
are linked to the quantum well exciton and cavity operators
$b_{\bf k}$ and $a_{\bf k }$ by an unitary Hopfield
transformation, namely,
\begin{equation}
\left (
\begin{array}{c}
b_{\bf k } \\
a_{\bf k }
\end{array}
\right ) = \left (
\begin{array}{cc}
M_{1,1,\bf k } & M_{1,2,\bf k} \\
M_{2,1,\bf k } & M_{2,2,\bf k} \\
\end{array}
\right )
\left (
\begin{array}{c}
p_{1,\bf k } \\
p_{2, \bf k }
\end{array}
\right )~.
\end{equation}
The matrix of Hopfield coefficients $M_{i,j,\bf k }$ is such that
$M_{1,1,\bf k } = M_{2,2,\bf k } = 1/\sqrt{1 + \rho_k^2}$ and
$M_{1,2,\bf k } = - M_{2,1,\bf k } = \sqrt{1-M_{1,1,\bf k }^2}$,
where $\rho_k = \hbar \Omega_R/(E_1(k)-E_C(k))$ and $2\hbar
\Omega_R$ is the polariton splitting when exciton and photon modes
are exactly resonant. Polaritons are interacting bosons, due to
the exciton-exciton exchange interaction and due to the anharmonic
part of the exciton-photon interaction (saturation)
\cite{Tassone_99,ciuti2000}, whose respective Hamiltonian
contributions $H_{XX}$ and $H^{sat}_{XC}$ are
\begin{equation}
H_{XX} = \frac{1}{2} \sum
\frac{\lambda_x^2}{A}\frac{6e^2}{\epsilon
\lambda_x}~b^{\dagger}_{{\bf k}+{\bf q}} b^{\dagger}_{{\bf
k'}-{\bf q}} b_{{\bf k}} b_{{\bf k'}}~,
\end{equation}
\begin{equation}
H^{sat}_{XC} =  - \sum \frac{\hbar \Omega_R}{n_{sat}
A}~a^{\dagger}_{{\bf k}+{\bf q}} b^{\dagger}_{{\bf k'}-{\bf q}}
b_{{\bf k}} b_{{\bf k'}} + \text{h.c.}~,
\end{equation}
being $A$ the excitation area, ${\lambda_X}$ the 2D exciton
radius, $\epsilon$ the static dielectric constant of the
semiconductor and $n_{sat} = 7/(16\pi \lambda_X^2)$ the exciton
saturation density.
 In the polariton basis, both effects contribute to create an
effective pair interaction potential. In our previous treatment of
polariton parametric scattering\cite{ciuti2000,Ciuti_01}, we
limited our decription to the lower branch. Including also the
upper branch, we get the following effective Hamiltonian
describing polariton-polariton interactions
\begin{equation}
H_{PP} = \frac{1}{2} \sum
\frac{\lambda_x^2}{A}V^{j_1,j_2,j_3,j_4}_{{\bf k},{\bf k'},{\bf
q}}~p^{\dagger}_{j_1,{\bf k}+{\bf q}} p^{\dagger}_{j_2,{\bf
k'}-{\bf q}} p_{j_3,{\bf k}} p_{j_4,{\bf k'}}~,
\end{equation}
where the effective branch-dependent potential is
\begin{eqnarray}
\label{potential}
 V^{j_1,j_2,j_3,j_4}_{{\bf k},{\bf k'},{\bf q}} =
 \left \{ \frac{6e^2}{\epsilon \lambda_x}
M_{1,j_1,{\bf k}+{\bf q}} M_{1,j_2,{\bf k'}-{\bf q}} M_{1,j_3,{\bf
k}} M_{1,j_4,{\bf k'}} \right .
\nonumber \\
- \frac{2\hbar\Omega_R}{n_{sat} \lambda_X^2} M_{2,j_1,{\bf k}+{\bf
q}} M_{1,j_2,{\bf k'}-{\bf q}} M_{1,j_3,{\bf k}} M_{1,j_4,{\bf
k'}} \nonumber \\
\left . - \frac{2\hbar\Omega_R}{n_{sat} \lambda_X^2} M_{2,j_4,{\bf
k'}} M_{1,j_3,{\bf k}} M_{1,j_2,{\bf k'}-{\bf q}} M_{1,j_1,{\bf k}
+ {\bf q}} \right \} ~,
\end{eqnarray}

Note that this Hamiltonian is for co-circularly polarized
polariton states. The first contribution to
$V^{j_1,j_2,j_3,j_4}_{{\bf k'},{\bf k'},{\bf q}}$ is proportional
to the 2D exciton binding energy $E_{\text{b}} = e^2/(2 \epsilon
\lambda_x)$ and is due to the exciton-exciton interaction. This
contribution is always repulsive, because $M_{1,j,{\bf k}}$ is
always positive. The other contribution is due to the anharmonic
exciton-photon coupling and can be either positive or negative,
depending on the branch indexes.

The regime of polariton parametric scattering takes place when a
pump laser drives coherently a single branch at a given
wave-vector. In this case, the corresponding quantum destruction
operator $p_{j_p,{\bf k_p}}$ can be approximated by the its
mean-field value $\langle p_{j_p,{\bf k_p}} \rangle$, which is a
classical field. Hence, the pair interaction Hamiltonian $H_{PP}$
can be approximated by the parametric Hamiltonian
\begin{equation}
H_{\text{par}} =  \sum_{j_1,j_2}\sum_{{\bf k}}
E^{j_1,j_2,j_p}_{{\bf k},{\bf k_p}} ~{\mathcal P}_{j_p,{\bf
k_p}}^2 ~p^{\dagger}_{j_1,{\bf k}} p^{\dagger}_{j_2,{2 \bf
k_p}-{\bf k}} + \text{h.c.}~,
\end{equation}
with
\begin{equation}
E^{j_1,j_2,j_p}_{{\bf k},{\bf k_p}} = ( V^{j_1,j_2,j_p,j_p}_{{\bf
k_p},{\bf k_p},{\bf k}-{\bf k_p}} + V^{j_2,j_1,j_p,j_p}_{{\bf
k_p},{\bf k_p},{\bf k}-{\bf k_p}} )/2~.
\end{equation}
The dimensionless pump polariton density is defined as $|{\mathcal
P}_{j_2,{\bf k_p}}|^2 = |\langle p_{j_p,{\bf
k_p}}\rangle|^2~\lambda_X^2/A$. The other effect is a mean-field
shift of the branch-dependent energy, namely $ \tilde{E}_j({\bf
k)} = E_j(k) + \Lambda^{j,j_p}_{{\bf k},{\bf k_p}}~ |{\mathcal
P}_{j_p,{\bf k_p}}|^2$, where $ \Lambda^{j,j_p}_{{\bf k},{\bf
k_p}} = ( V^{j,j_p,j,j_p}_{{\bf k},{\bf k_p},{\bf 0}} +
V^{j_p,j,j_p,j}_{{\bf k_p},{\bf k},{\bf 0}}+ V^{j_p,j,j,j_p}_{{\bf
k},{\bf k_p},{\bf k_p}-{\bf k}} + V^{j,j_p,j_p,j}_{{\bf k_p},{\bf
k},{\bf k}-{\bf k_p}})/2$.

\begin{figure}
\includegraphics[width=7cm]{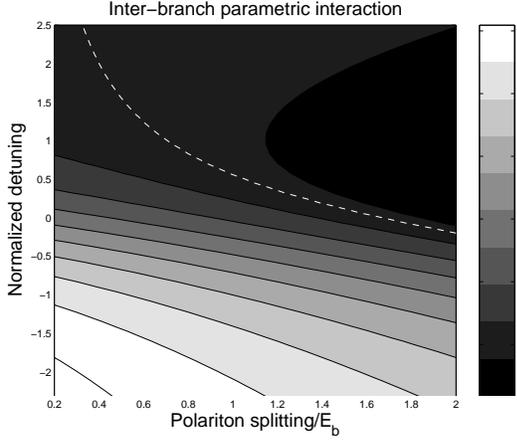}
\caption{(a) Contours of the dimensionless parametric interaction
energy $E^{1,2,2}_{k_{\text{r}},0}/E_{\text{b}}$ versus $2\hbar
\Omega_R/E_{\text{b}}$ and the normalized detuning $\delta =
(E_C(0)-E_X)/(2\hbar \Omega_R)$. Parameters: $E_C(0) =
1.5~e\text{V}$, exciton binding energy $E_{\text{b}} =
10~\text{m}e\text{V}$. The white-dashed line depicts the zero
value points. \label{E_int_circle}}
\end{figure}

In this Section, we are interested in the case of pump excitation
of the upper branch ($j_p = 2$), with the final states belonging
to two different branches ($j_1 \neq j_2$). Since $E^{1,2,2}_{{\bf
k},{\bf k_p}} = E^{2,1,2}_{{\bf k},{\bf k_p}}$,
 the parametric
interaction Hamiltonian reads
\begin{equation}
H_{\text{par}} =  \sum_{{\bf k}} E^{1,2,2}_{{\bf k},{\bf k_p}}
~{\mathcal P}_{2,{\bf k_p}}^2 (p^{\dagger}_{1,{\bf k}}
p^{\dagger}_{2,2{\bf k_p}-{\bf k}} + p^{\dagger}_{2,{\bf k}}
p^{\dagger}_{1,2{\bf k_p}-{\bf k}})+ \text{h.c.}~. \label{H}
\end{equation}
When applied on the vacuum state $|0\rangle$, $H_{\text{par}}$
generate pairs of polaritons with total in-plane momentum $2{\bf
k_p}$, which are entangled with respect to the branch index.
Indeed, Eq. (\ref{H}) has the paradigmatic form of Hamiltonian,
describing the generation of EPR pairs of bosons, which are
entangled with respect to a discrete degree of freedom. In quantum
optics, the literature about the non-classical photon properties
associated to this Hamiltonian is impressive\cite{Shih}. In our
case, entanglement concerns polaritonic particles and one peculiar
polaritonic degree of freedom, namely the branch index. The
generation of branch-entangled pairs is allowed only when there is
phase-matching for the two branch-exchanged processes, i.e.
$E_1({\bf k})+E_2(2{\bf k_p}-{\bf k}) = 2E_2({\bf k_p})$ and
$E_2({\bf k})+E_1(2{\bf k_p}-{\bf k}) = 2E_2({\bf k_p})$. For
${\bf k_p} \neq {\bf 0}$, there are only two possible signal and
idler wave-vectors, which are the intersection points in Fig.
\ref{fig_circles}(b), as anticipated. When ${\bf k_p} = {\bf 0}$,
branch-entanglement is achievable for every pair of in-plane
wave-vectors $({\bf k},-{\bf k})$ on the phase-matching ring
$|{\bf k}| = k_{\text{r}}$.  Fig. \ref{E_int_circle} shows the
contours of the interaction energy $E^{1,2,2}_{{\bf
k_{\text{r}}},{0}}$ (units of the exciton binding energy
$E_{\text{b}}$), as a function of the polariton splitting to
binding energy ratio $2\hbar \Omega_R/E_{\text{b}}$ and of the
normalized detuning $\delta = (E_C(0)-E_X)/(2\hbar \Omega_R)$. As
anticipated, Fig. \ref{E_int_circle} shows that the effective
interaction can be either positive or negative (the change of sign
occurs across the white-dashed line). The effective interaction is
positive when it is dominated by the exciton-exciton interaction,
negative when the anharmonic exciton-photon coupling takes over.
\subsection{Emission of frequency-entangled photon pairs}
\label{out} The intra-cavity polariton parametric scattering
dynamics is coupled to the extra-cavity field, giving rise to
parametric luminescence\cite{Ciuti_01}. This coupling is usually
described by the quasi-mode Hamiltonian
\begin{equation}
H_{ext} = \sum_{j,{\bf k}} \int d\omega ~g(\omega)~ |M_{j,2,{\bf
k}}|^2~\alpha^{\dagger}_{\omega,{\bf k}}~ p_{j,{\bf k}}
 + \text{h.c.}~,
\end{equation}
where $g(\omega)$ is the coupling energy (approximately constant
in the mirror spectral stop-band) and
$\alpha^{\dagger}_{\omega,{\bf k}}$ is the creation operator of an
extra-cavity photon with energy $\hbar \omega$ and conserved
in-plane wave-vector ${\bf k}$. The free space photon is emitted
with an external angle $\theta$ with respect to the vertical
direction, such as $k = (\omega/c) \sin \theta$. The coupling of
each branch ($j \in \{1,2\}$) to the external field is
proportional to the photonic fraction $|M_{j,2,{\bf k}}|^2$.
\begin{figure}[t!]
\includegraphics[width=8cm]{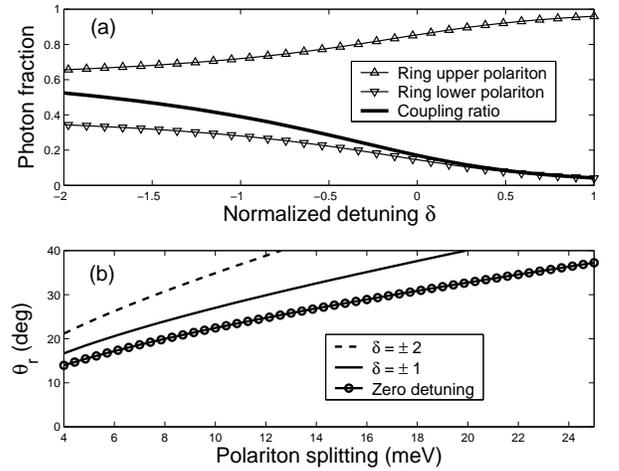}
\caption{ (a) Photon fractions  of the polariton modes on the
phase-matching ring ($|{\bf k}| = k_{\text{r}}$) as a function of
the normalized detuning . Upper triangle: upper branch. Lower
triangle: lower branch. Thick solid line: the ratio between the
lower and upper branch photon fractions. (b) Ring emission
external angle $\theta_{\text{r}}$ (deg) versus polariton
splitting (meV) for different normalized detunings.
 \label{fig4}}
\end{figure}
Importantly, branch-entangled pairs of polaritons can emit
frequency-entangled pairs of photons, i.e. states like
\begin{equation}
|\Psi \rangle \propto
 ( \alpha^{\dagger}_{\omega_1,{\bf k_{\text{r}}}}
   \alpha^{\dagger}_{\omega_2,-{\bf k_{\text{r}}}}
 +~
  \alpha^{\dagger}_{\omega_2,{\bf k_{\text{r}}}}
  \alpha^{\dagger}_{\omega_1,-{\bf k_{\text{r}}}}
 )  |0\rangle~,
\end{equation}
where $\hbar \omega_1$ ($\hbar \omega_2$) is
 the energy of the lower (upper) branch state
 with in-plane wave-vector $k_{\text{r}}$.
The frequency-entanglement\cite{note} of photon pairs can be
measured by coincidence counting in Hong-Ou-Mandel-type
interferometers\cite{interferometer}, which are also used in
quantum tomography\cite{tomography}. In order to have a
significant extra-cavity visibility, the polariton signal and
idler modes need to have a similar coupling to the extra-cavity
field. This occurs when the cavity photon fraction of the
polariton signal and idler modes is comparable. Fig. \ref{fig4}(a)
depicts respectively the photon fractions $|M_{2,2, k_{\text{r}}
}|^2$ and $|M_{1,2,k_\text{r}}|^2$ of the upper and lower branch
modes on the ring, versus the normalized detuning. The thick solid
line shows the ratio $|M_{1,2,k_\text{r}}/M_{2,2,k_\text{r}}|^2$.
Compared to the known intra-branch process\cite{Savvidis_00} where
the signal-idler coupling ratio is typically less than $0.05$
\cite{twin}, the inter-branch process here described enjoys a
higher ratio. At zero detuning, the ratio is $\simeq 0.2$, rising
significantly in the region of negative detuning ($\simeq 0.4$ for
$\delta = -1$). Finally, Fig. \ref{fig4}(b) shows the dependence
of the phase-matching ring wave-vector on the polariton coupling.
The corresponding emission angle $\theta_{\text{r}}$ (deg)
increases with increasing polariton splitting. For a given
polariton splitting, $\theta_{\text{r}}$ depends only on
$|\delta|$, being minimum for zero detuning.

\subsection{Losses for the polariton modes}
As well known in quantum optics, the interesting quantum regime is
achieved when the scattering is spontaneous, i.e. the probability
of having more than one entangled pair in the same state is
negligible. In other words, the parametric scattering should be
kept below the stimulated parametric oscillation
threshold\cite{Ciuti_01,Whittaker}. However, the system can not be
driven too much below threshold, because other scattering
mechanism can prevail, disentangling the pairs created by
parametric scattering. Hence, the role of losses is crucial and
needs to be carefully addressed.
\begin{figure}
\includegraphics[width=7cm]{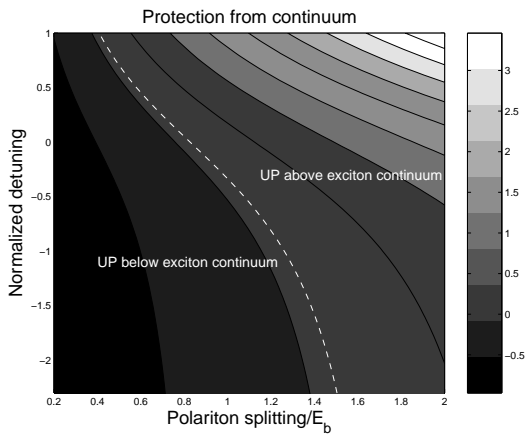}
\caption{Contours of
$(E_2(k_{\text{r}})-E_X-E_{\text{b}})/E_{\text{b}}$ versus $2\hbar
\Omega_R/E_{\text{b}}$ and the normalized detuning $\delta =
(E_C(0)-E_X)/(2\hbar \Omega_R)$. The white-dashed line depicts the
zero value points, i.e. the upper polariton state on the ring is
resonant with the continuum band-edge. Same parameters as in Fig.
\ref{E_int_circle}. \label{E_int_continuum}}
\end{figure}

\label{losses}

\paragraph{Linear losses}
Losses for the polariton modes produce a branch- and wave-vector
dependent polariton broadening $\gamma_{j,k}$. In the low
excitation regime at low temperatures, the linear broadening
$\gamma^{L}_{j,k}$ is essentially due to the radiative linewidth,
the interaction with acoustic phonons \cite{tassone}, scattering
by impurities and, for the upper branch, mixing with the exciton
continuum states\cite{citrin}. The radiative lifetime and the
impurity concentration are strongly sample-dependent, being
determined by the growth quality of the microcavity. Usually, the
broadening due to emission of acoustic phonons is smaller with
respect to the radiative linewidth and to the impurity-induced
losses. On the other hand, the continuum of unbound electron-hole
pairs is a major source of broadening for the upper branch states
with energy higher than the continuum onset. In principle, the
upper branch state on the ring can form a Fano resonance with the
continuum states, with a finite probability of decaying
irreversibly into undesirable unbound electron-hole pairs. This
issue is addressed in Fig. \ref{E_int_continuum}, which shows the
difference between the upper branch final-state energy
$E_2({k_\text{r}})$ and the continuum band-edge energy
$E_X+E_{\text{b}}$, in units of $E_{\text{b}}$. The white-dashed
line depicts the points where the difference is 0. The encouraging
fact is that there is a wide region with negative values, implying
that the upper polariton final-state can be protected from the
free carrier absorption. At zero detuning, this occurs for a
polariton splitting to exciton binding energy ratio smaller than
$0.8$. The condition becomes even less stringent for negative
detunings.

\paragraph{Density-dependent losses}
\begin{figure}[t]
\includegraphics[width=8cm]{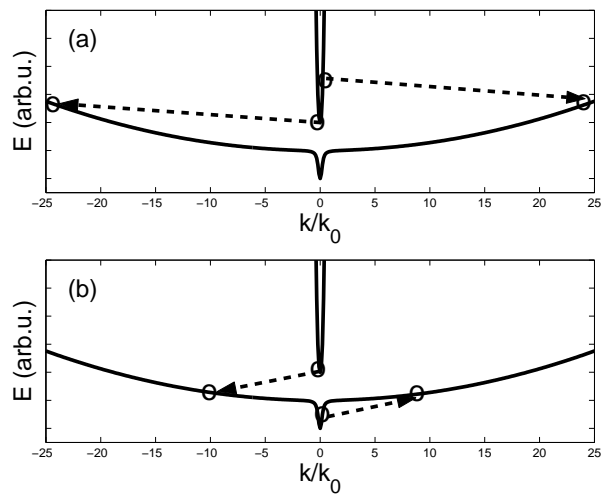}
\caption{Sketch of the pair scattering processes responsible for
nonlinear losses of the polariton modes. (a) An upper branch
polariton scatter with one pump polariton ($k_{p} = 0$) belonging
to the upper branch. The final states are excitons with large
momentum. (b) Analogous loss process for a lower polariton, due to
scattering with one pump polariton. \label{reservoir}}
\end{figure}
For moderate and higher excitation densities, nonlinear
losses\cite{baumberg_98,Ciuti_thr,Savasta_03,Baars,Huynh_PRB} play
an important role. In particular, polariton pair scattering into
the exciton reservoir can become the leading source of broadening
for the polariton modes. Namely, fast decoherence of the pumped
mode can occur due to pump-pump scattering into the high-momentum
exciton states, while pump-signal (idler) scattering into the
exciton reservoir creates a loss mechanism for the polariton
signal (idler) mode. Panel (a) represents the scattering of one
upper polariton state with one pump polariton with zero in-plane
wave-vector. Panel (b) represents the analogous scattering for one
lower polariton. Within the Born approximation, the nonlinear
broadening is given by

\begin{equation}
\gamma^{NL}_{j,k} = 2\pi \sum_{{\bf q}} N_{2,{\bf 0}} \left |
(\lambda_X^2/A) V^{1,1,2,j}_{{\bf 0},{\bf k},{\bf q}} \right |^2
\delta (\Delta E)~,
\end{equation}
where here $\delta$ is the Dirac function, $\Delta E = E_{2}(0)+
E_{j}(k) - E_1(q) - E_1(|-{\bf q}+{\bf k}|)$ and $N_{2,{\bf 0}}$
is the number of polaritons in the pumped mode. If the pump mode
is driven coherently (the case of our interest), $N_{2,{\bf 0}}
\simeq |\langle p_{2,{\bf 0}} \rangle|^2$.
 Since the energy conservation is fulfilled for a
wave-vector $q$ very large compared to $k$ (see
Fig.\ref{reservoir}), we can safely approximate $E_1(|-{\bf
q}+{\bf k}| ) \simeq E_1(q) \simeq E_X + \frac{\hbar^2 q^2}{2 M}$,
being $M$ the exciton mass. Hence, the expression for the
nonlinear broadening becomes
\begin{equation}
\gamma^{NL}_{j,k} \simeq \frac{M\lambda_X^2}{2\hbar^2} |
V^{1,1,2,j}_{{\bf 0},{\bf k},\bar{q}_j}|^2~(n_{2,{\bf
0}}\lambda_X^2),
\end{equation}
where $\bar{q}_j$ is such that $E_{2}(0)+ E_{j}(k) = 2
E_1(\bar{q}_j)$ and $n_{2,{\bf 0}} = N_{2,{\bf 0}}/A$ is the
density of pump polaritons per unit area. Let us calculate the
nonlinear broadening for a set of realistic parameters, namely
exciton mass $M=0.3~m_0$, pump density $n_{2,{\bf 0}} =
\frac{1}{20}~n_{sat}$, polariton splitting $2\hbar \Omega_R =
7~\text{m}e\text{V}$, $\lambda_X = 10~\text{nm}$. For this
parameters, we get $\gamma^{NL}_{1,k_{\text{r}}} =
1.1~\text{m}e\text{V}$, $\gamma^{NL}_{2,k_{\text{r}}} =
0.25~\text{m}e\text{V}$ for normalized photon detuning $\delta = +
1$. For $\delta = 0$, $\gamma^{NL}_{1,k_{\text{r}}} =
4.3~\text{m}e\text{V}$, $\gamma^{NL}_{2,k_{\text{r}}} =
0.3~\text{m}e\text{V}$, while for $\delta =  - 0.5$
$\gamma^{NL}_{1,k_{\text{r}}} = 6.7~\text{m}e\text{V}$,
$\gamma^{NL}_{2,k_{\text{r}}} = 0.12~\text{m}e\text{V}$. Note
that, under pumping of the upper branch,
 the collision broadening of the upper polariton state on the
ring is smaller than that of the companion state on the lower
branch. This occurs because the upper polariton state on the ring
has always an excitonic fraction smaller than the lower polariton
state with the same wave-vector.

\subsection{Collision broadening catastrophe}
\label{threshold} The spontaneous scattering regime\cite{Ciuti_01}
is achieved for pump intensities well below the stimulated
parametric oscillation threshold. Since inter-branch parametric
interaction and pair scattering into the exciton reservoir are due
to the same microscopic mechanism, {\it a priori} it is not clear
if a stimulation threshold can be ever achieved under pump
excitation of the upper branch. In fact, the parametric
oscillation threshold is achieved when the parametric interaction
energy compensates for the total losses of the signal-idler pair,
namely
\begin{equation}
|E^{1,2,2}_{{\bf k_{\text{r}}},{0}} ~{\mathcal P}_{2,{\bf
0}}^{\text{thr}~ 2}|^2 = (\gamma^L_{1,k_{\text{r}}}+
\gamma^{NL}_{1,k_{\text{r}}}) (\gamma^L_{2,k_{\text{r}}}+
\gamma^{NL}_{2,k_{\text{r}}})~, \label{thr}
\end{equation}
which is a self-consistent equation, because $
\gamma^{NL}_{j,k_{\text{r}}}$ depends on the pump density. If we
define $\xi^{NL}_{j,k} = \frac{M\lambda_X^2}{2\hbar^2} |
V^{1,1,2,j}_{{\bf 0},{\bf k},\bar{q}_j}|^2$, then we can rewrite
the collision broadening as $\gamma^{NL}_{j,k} = \xi^{NL}_{j,k}
n_{2,{\bf 0}}\lambda_X^2$. Hence, Eq. (\ref{thr}) becomes
\begin{equation}
\left [(E^{1,2,2}_{{\bf k_{\text{r}}},{0}})^2 -
\xi^{NL}_{1,k_{\text{r}}} \xi^{NL}_{2,k_{\text{r}}}\right]
(n^{thr}_{2,{\bf 0}}\lambda_X^2 )^2 = \beta~n^{thr}_{2,{\bf
0}}\lambda_X^2 + \gamma^L_{1,k_{\text{r}}}
\gamma^L_{2,k_{\text{r}}}~,
\end{equation}
where $\beta = (\gamma^{L}_{1,k_{\text{r}}}
\xi^{NL}_{2,k_{\text{r}}}+ \gamma^{L}_{2,k_{\text{r}}}
\xi^{NL}_{1,k_{\text{r}}})$ is always positive. For typical values
of the exciton mass, the quantity $(E^{1,2,2}_{{\bf
k_{\text{r}}},{0}})^2 - \xi^{NL}_{1,k_{\text{r}}}
\xi^{NL}_{2,k_{\text{r}}}$ is negative. Hence, Eq. (\ref{thr}) can
be never satisfied, because the left-hand side is negative, while
the right-hand side is always strictly positive. In other words,
the collision broadening due to scattering into the high-momentum
states acts as a positive feed-back, preventing the system to
enter the stimulated regime. This kind of collision catastrophe is
absent when the pump excites the lower branch, because the
coupling to the high-momentum states is strongly suppressed
\cite{baumberg_98,Ciuti_thr,Baars,Savasta_03,Huynh_PRB}.
\begin{figure}[t!]
\includegraphics[width=7.8 cm]{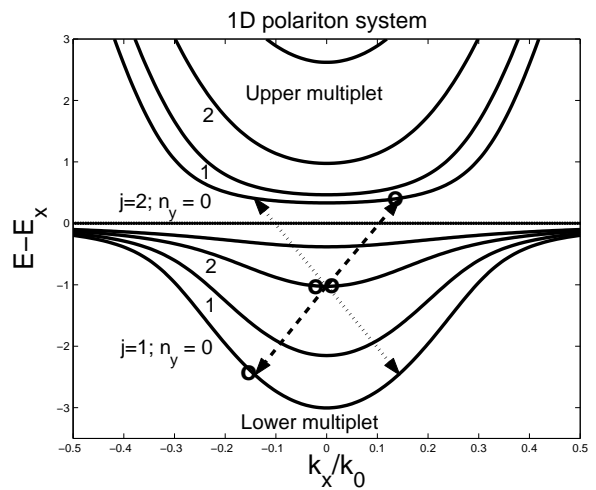}
\caption{Energy dispersion (units of $2 \hbar \Omega_R$) of
1D-polaritons as a function on the wave-vector $k_x$ ($k_0$ units)
along the direction of the photonic wire. Compared to the
2D-system, the lower branch ($j=1$) is split in a multiplet of
sub-branches ($n_y = 0,1,2,..$), as well as the upper branch ($j =
2$). The arrows depict the considered inter-branch parametric
scattering process, in which the pump excites the $n_y = 2$ lower
sub-branch mode with $k_x = 0$ . Parameters: $2\hbar \Omega _R =
4~\text{m}e\text{V}$, wire width $L_y = 4~\mu\text{m}$,
 $E_X = E_C(0) + 4\hbar \Omega_R$, with $E_C(0) = 1.5~e\text{V}$ is the 2D-cavity
 energy.\label{1D_scattering}}
\end{figure}

\section{1D microcavities}
\label{wire}

\begin{figure}[t!]
\includegraphics[width=8cm]{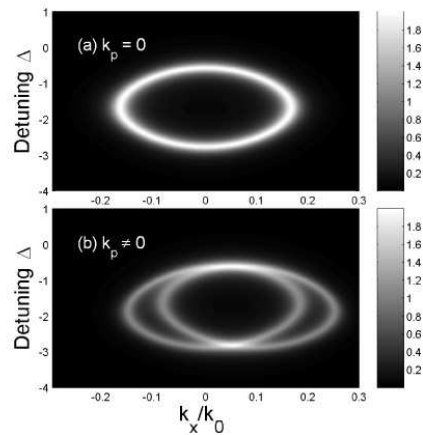}
\caption{Phase-matching function for the inter-subbranch
scattering of 1D-polaritons depicted in Fig. \ref{1D_scattering},
as a function of $k_x$ ($k_0$ units) and the normalized detuning
$\Delta = (E_C(0) - E_X)/(2 \hbar \Omega)$, with $E_C(0)$ the
2D-cavity energy.  (a) ${\bf k}_p = {\bf 0}$. (b) ${\bf
k_{\text{p}}} = 0.05~{k_0}~\hat{\bf x}$. \label{ellipses_1D}.
Parameters: $2\hbar \Omega _R = 4~\text{m}e\text{V}$, $\gamma =
0.5$~\text{m}e\text{V}}
\end{figure}
The concept of branch entanglement is quite general and can be
applied also to multi-branch systems, such as photonic wires
\cite{Dasbach_wire}. In a one-dimensional cavity, the additional
confinement along the y-direction produces a series of cavity
photon sub-branches, whose energy dispersion $E^{(n_y)}_C(k_x)$ is
given by
\begin{equation}
\left ( E^{(n_y)}_C(k_x)\right )^2 = \left ( E_C(k_x) \right )^2 +
\frac{(\hbar c)^2}{\epsilon} \left (\frac{\pi (n_y+1)}{L_y}\right
)^2
\end{equation}
where $E_C(k_x)$ is the energy of the planar cavity with $k =
k_x$, $L_y$ is the wire width and $n_y$ is the sub-branch index
(positive or equal to zero). Strong coupling to the exciton
resonance produces a many-fold of lower polariton sub-branches
with energy $E^{(n_y)}_1(k_x)$ and upper polariton sub-branches
with energy $E^{(n_y)}_2(k_x)$. Each cavity sub-band couples to an
exciton mode with the same symmetry\cite{Tarta_wire}. The
polariton splitting $2\hbar \Omega_R$ is approximately
independent\cite{Tarta_wire} of the branch index $n_y$ for small
values of $n_y$.
\begin{figure}[t!]
\includegraphics[width=7.8 cm]{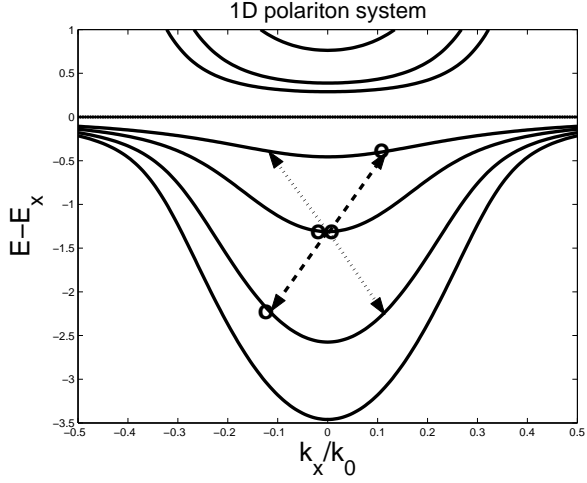}
\caption{Arrows depict the inter-sub-branch scattering, with all
states belonging to the lower many-fold. Parameters: $2\hbar
\Omega _R = 4~\text{m}e\text{V}$, wire width $L_y = 4
\mu\text{m}$,
 $E_X = E_C(0) + 3.5~\hbar \Omega_R$, with $E_C(0) = 1.5~e\text{V}$ is the 2D-cavity
 energy. Horizontal and vertical axis as in Fig.
 \ref{1D_scattering}.
 \label{1D_scattering_onlyLP}}
\end{figure}
As experimentally demonstrated in the experiments by Dasbach {\it
al.}\cite{Dasbach_wire}, there are many new parametric scattering
channels available. In particular, it is possible to have
inter-branch scattering by pumping one lower
sub-branch\cite{Dasbach_wire}. The momentum conservation along the
y-direction is lifted, being replaced by the less stringent {\it
parity} selection rule\cite{Dasbach_wire,Dasbach_thesis}. This
selection rule for pair scattering of 1D-polaritons imposes that
the sum of $n_y$ for signal and idler must be even. The
inter-branch parametric scattering process has an
efficiency\cite{Woods}, which is comparable to the intra-branch
scattering in planar microcavities.
 In Fig.\ref{1D_scattering}, we
propose a scattering process, in which the pump excites the lower
sub-branch with $n_y = 2$ and $k_x = 0$. For a proper
exciton-photon detuning, there is a phase-matched process, in
which the final states are two polariton modes with opposite and
finite wave-vectors, one belonging to the lower $n_y = 0$
sub-branch and the other to the upper $n_y = 0$ sub-branch. The
phase-matching function for this inter-branch scattering channel
is depicted in Fig. \ref{ellipses_1D}, as a function of the
wave-vector $k_x$ and the normalized detuning $\Delta = (E_C(0) -
E_X)/(2 \hbar \Omega)$, where $E_C(0)$ is the 2D cavity energy and
$2 \hbar \Omega_R$ is the polariton splitting. As in Fig.
\ref{fig_circles}, the phase-matching function is equal to $2$,
when there are two branch-exchanged processes, which are exactly
phase-matched (the condition for branch-entanglement). For zero
pump wave-vector $k_p$ (see Fig.\ref{ellipses_1D}(a)), this
property is achieved in a broad, but finite range of negative
detuning $\Delta$. In contrast to the 2D-case, for $k_p \neq 0$,
the phase-matching function is equal to 2 only at the pump
wave-vector, as shown in Fig.\ref{ellipses_1D}(b). But this does
not correspond to pure polariton branch-entanglement, because
signal and idler have the same wave-vector.

\begin{figure}[t!]
\includegraphics[width=8cm]{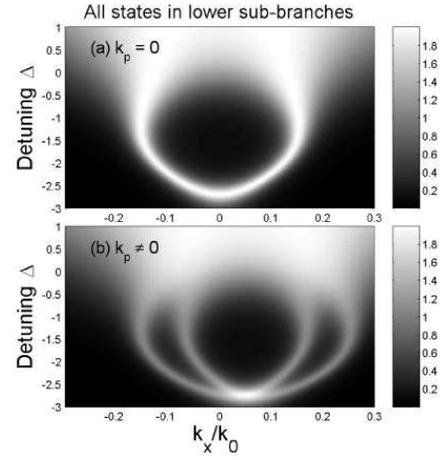}
\caption{Phase-matching function for the inter-subbranch
scattering of 1D-polaritons depicted in Fig.
\ref{1D_scattering_onlyLP}, as a function of $k_x$ ($k_0$ units)
and the normalized detuning $\Delta = (E_C(0) - E_X)/(2 \hbar
\Omega)$, with $E_C(0)$ the 2D-cavity energy.  (a) ${\bf k}_p =
{\bf 0}$. (b) ${\bf k_{\text{p}}} = 0.05~{k_0}~\hat{\bf x}$.
\label{phase_matching_onlyLP_1D}. Parameters: $2\hbar \Omega _R =
4~\text{m}e\text{V}$, $\gamma = 0.4$~\text{m}e\text{V}}
\end{figure}

Importantly, in a photonic wire it is possible to have
inter-sub-branch scattering processes restricted to the lower
many-fold only. One parity-conserving process is shown in Fig.
\ref{1D_scattering_onlyLP}, where the pump excites the $n_y = 2$
sub-branch at $k_x = 0$ and the signal and idler modes belong to
the $n_y = 1$ and $n_y = 3$ sub-branches. The phase-matching
properties of this processes are reported in
Fig.\ref{phase_matching_onlyLP_1D} as a function of the signal
wave-vector along the wire direction and of the normalized
detuning $\Delta$.

The interest of photonic wires does not rely only in the
possibility of having new scattering channels. One advantage is to
provide a much better protection from the exciton reservoir. In
fact, in contrast to pumping of the upper branch , the
inter-branch process shown in Fig.\ref{1D_scattering} suffers much
weaker nonlinear losses due to pair scattering into the
high-momentum exciton states. As already studied theoretically and
experimentally, under excitation of the lower branch, pump-pump
scattering into the exciton reservoir is strongly suppressed due
to lack of energy-momentum
conservation\cite{Ciuti_thr,Baars,Huynh_PRB}. The same is true for
pump-signal and pump-idler scattering. The only allowed channel is
the signal-signal (or idler-idler) scattering, in which the signal
(idler) mode belong to the upper branch. But this is not a crucial
process especially below or near threshold, when the signal
(idler) population is much smaller than the pump one.

\section{Conclusions}

\label{conclusion} In conclusion, we have proposed and analyzed a
scheme for the generation of branch-entangled polariton pairs in
semiconductor microcavities through spontaneous inter-branch
parametric scattering. Branch-entanglement of polariton pairs
leads to emission of frequency-entangled pairs of extra-cavity
photons, which have been recently attracting considerable
attention in the field of quantum tomography\cite{tomography}.
This kind of non-classical states can not be achieved by
intra-branch polariton pair scattering\cite{Savvidis_00}, being a
peculiarity of inter-branch processes. In planar microcavities,
the phase-matching conditions are satisfied by pumping the upper
polariton branch for an arbitrary pump in-plane wave-vector ${\bf
k_{\text{p}}}$. We have studied the phase-matching properties and
the efficiency of the process as a function of exciton-photon
detuning, polariton splitting and exciton binding energy. While
the phase-matching properties for the 2D inter-branch process are
very flexible, the nonlinear losses due to polariton pair
scattering into the high-momentum exciton states is a reason of
concern, being a significant source of decoherence.  The lack of
protection of pump polaritons in the upper branch  can be
naturally overcome in photonic wires, thanks to the existence of a
many-fold of sub-branches. In this paper, we have shown that there
are parity-conserving inter-branch scattering processes (forbidden
in planar microcavities), in which the pump excites a lower
polariton sub-branch mode with $k_x =0$, providing
branch-entanglement of the signal-idler polariton pair. These
processes enjoy much better protection from the high-momentum
exciton states, making one-dimensional microcavities a strong
candidate to demonstrate and exploit the quantum effects here
proposed. We hope that the ideas presented in this paper will
stimulate experimental and theoretical research in a field at the
frontier between condensed matter physics and quantum optics.
Indeed, one intriguing goal would be the development of polariton
micro-sources of non-classical states with controllable
properties.

We wish to thank J. Tignon , G. Bastard, G. Dasbach, Ph.
Roussignol, M. Saba for discussions. LPA-ENS (former LPMC-ENS) is
"Unit\'{e} Mixte de Recherche Associ\'{e} au CNRS (UMR 8551) et
aux Universit\'{e}s Paris 6 et 7".


\begin{thebibliography}{}
\bibitem{note} The emission angles corresponding to
$\omega_1$ and $\omega_2$ are slightly different ($\Delta
\theta_{\text{r}}/\theta_{\text{r}} \simeq
(\omega_2-\omega_1)/\omega_1$, typically less than $1\%$). Hence,
the emitted photon pair is also momentum-entangled.
\end{thebibliography}
\end{document}